# Ferromagnetic and superconducting instabilities in graphite

Y. Kopelevich, S. Moehlecke, and R. R. da Silva

Instituto de Física "Gleb Wataghin", Universidade Estadual de Campinas, Unicamp 13083-970, Campinas, São Paulo, Brasil

## 1. INTRODUCTION

A few years ago two of us reported on anomalous behavior of graphite in magnetic field [1, 2]. In particular, the metal-insulator transition driven by the magnetic field applied perpendicular to graphite basal planes [1], as well as ferromagnetic- and superconducting-like hysteresis loops [2] have been revealed in highly oriented pyrolitic graphite (HOPG) samples at room temperature (for review articles see [3, 4]).

Shortly after, room temperature ferromagnetism has been observed in polymerized rhombohedral (rh) $C_{60}$ samples [5-7], extraterrestrial graphite [8], and materials consisting of curved graphite-like sheets such as glassy and microporous carbon [9, 10].

The accumulated so far experimental evidences indicate that a structural disorder, topological defects, as well as adsorbed foreign atoms can be responsible for the occurrence of both ferromagnetic and superconducting patches in graphitic structures, i. e. $sp^2$-bonded carbon atoms. In particular, it has been demonstrated that sulfur atoms adsorption triggers superconducting instabilities in graphite [11-14], whereas proton irradiation [15] enhances the sample ferromagnetic response.

In the present Chapter we review our most recent work related to the studies of magnetic properties of graphite and related carbon materials.

The Chapter is organized as follows.

In section 2 we present a comparative analysis of ferromagnetic properties of microporous carbon, HOPG, and rh-$C_{60}$ samples. Results demonstrating an oxygen adsorption effect on the magnetic properties of graphite are presented in section 3. Section 4 is devoted to the ferromagnetic and superconducting properties of graphite-sulfur composites. Discussion of the results and concluding remarks are given in section 5.

## 2. FERROMAGNETISM IN CARBON MATERIALS: A COMPARATIVE STUDY

It appears that the room-temperature ferromagnetism in rh-$C_{60}$ compounds [5-7] occurs only in samples prepared very close to the temperature at which the fullerene cages collapse and graphitized carbon forms. In agreement with such observation are experiments performed on glassy carbon [9] which demonstrated that the ferromagnetism emerges during the graphitization process.

On the theoretical side, the occurrence of both ferromagnetic and superconducting instabilities due to topological disorder in graphitic sheets has been predicted [16]. The analysis given in Ref. [16] assumes the formation of pentagons and heptagons, i. e. disclinations in the graphene honeycomb lattice. The low-lying electronic states of an isolated graphitic sheet can be well approximated by the Dirac equations in (2+1) dimensions. Then, according to Ref. [16], a random distribution of topological defects described in terms of a random gauge field can lead to an enhancement of the density of states of Dirac fermions N(E) at low energies E, and hence to magnetic or superconducting instabilities. Compounds with curved graphene layers have been proposed as promising materials for both ferromagnetism and superconductivity occurrence [16].

These experimental as well as theoretical results motivate us to explore the magnetic behavior of microporous carbon (MCY), a three-dimensional nano-arrayed structure whose arrangement matches that of supercages of zeolite Y.

The microporous carbon samples have been prepared at the Tohoku University [17, 18] by the following template technique. Powder zeolite Y was impregnated with furfuryl alcohol (FA) and the FA was polymerized inside the zeolite channels by heating the FA/zeolite composite at 150 °C under $N_2$ flow. The resultant PFA (polyfurfuryl alcohol)/zeolite composite was heated to 700 °C in $N_2$. As soon as the temperature reached 700 °C, propylene chemical vapor deposition (CVD) was performed at this temperature to further deposit carbon. After the CVD, the composite was heat-treated again at 900 °C under a $N_2$ flow. The resultant carbon was liberated from the zeolite framework by acid washing. The obtained microporous carbon possesses a very high surface area of 3600 $m^2$/g and consists of curved 3D graphene networks which may contain randomly distributed pentagons and heptagons carbon rings. The microporous carbon particle size ranges from 1000 Å to 4000 Å. Spectrographic analysis of the MCY samples revealed magnetic impurity contents of Fe (64 ppm), Co (4.4 ppm), and Ni (3.5 ppm) [10].

Dc magnetization M(H,T) measurements were always performed using a SQUID magnetometer MPMS5, Quantum Design.

Figure 1 presents the low-field portions of the magnetization hysteresis loops M(H) measured for MCY, HOPG (Union Carbide Co.), and Rh-$C_{60}$ samples at T = 300 K demonstrating an enhanced ferromagnetic signal in MCY and Rh-$C_{60}$ samples as compared to that in HOPG.

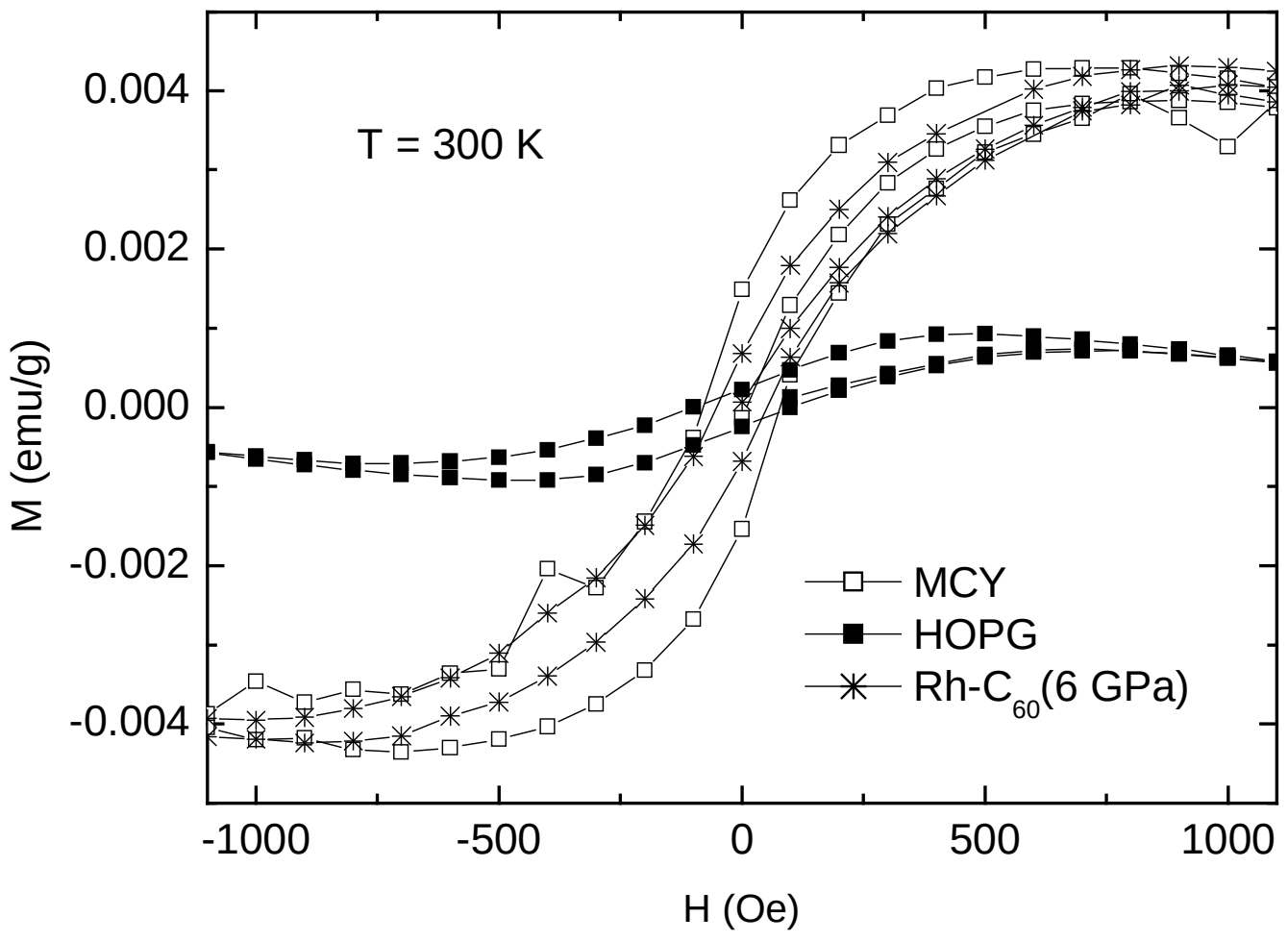

Figure 1. Magnetization hysteresis loops M(H) measured for MCY, HOPG, and Rh-$C_{60}$ samples. In the case of HOPG, the magnetic field was applied nearly parallel to the sample basal planes; because of small (≤ 3 °) field/planes misalignment, M(H) decreases at high enough fields due to Landau diamagnetism associated with the graphene planes.

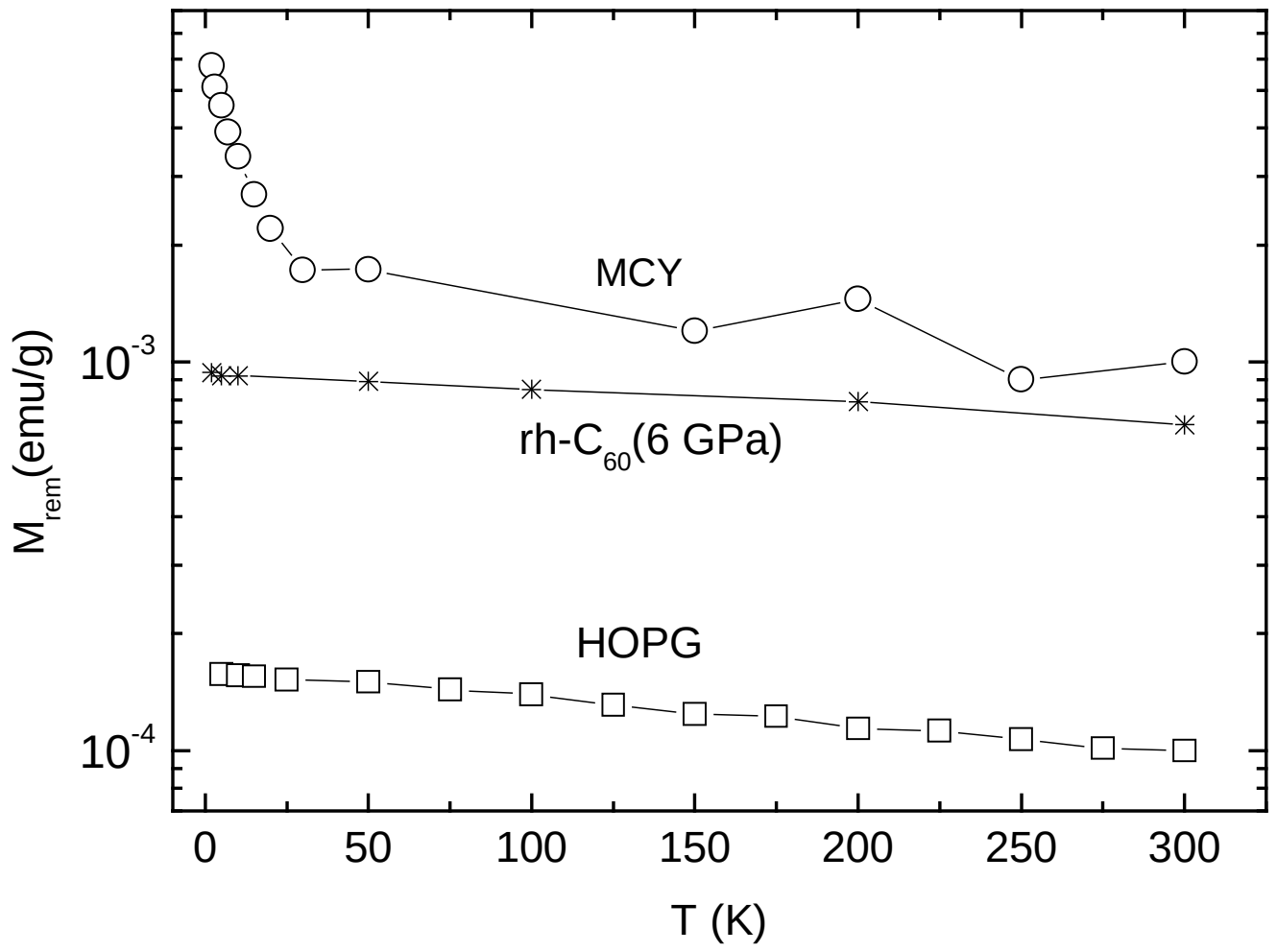

Figure 2. Remnant magnetization $M_{rem}$(T) obtained for MCY, HOPG, and rh-$C_{60}$ samples.

Because of the unavoidable in-plane diamagnetic contribution (due to intrinsic misalignment of the graphite layers with respect to each other occurring even in the best HOPG samples) we have chosen the remnant magnetization $M_{rem}(H = 0) = [M^+(H = 0) – M^-(H = 0)]/2$ for a comparative analysis. Shown in Fig. 2 is $M_{rem}(T)$ obtained for MCY, HOPG, and rh-$C_{60}$ samples, where $M^+(H = 0)$ and $M^-(H = 0)$ are zero-field positive and negative magnetizations measured after the field cycling. As Fig. 2 demonstrates, $M_{rem}$(MCY) >> $M_{rem}$(HOPG) for all temperatures, and is comparable to $M_{rem}(T)$ measured for rh-$C_{60}$ sample prepared at T = 1073 K and pressure P = 6 GPa [19]. We stress that the results of x-ray structural analysis performed on this nominal rh-$C_{60}$ sample revealed the coexistence of the rh-$C_{60}$ phase and clusters of graphite-like layers [19].

Also a comparative analysis of the data obtained on microporous carbon and rh-$C_{60}$ sample synthesized at the pressure 9 GPa and T = 800 K [6] reveals a striking correspondence between M(H) measured in these materials. As follows from Fig. 3, the remnant magnetizations measured for MCY and rh-$C_{60}$ (9 GPa) samples practically coincide.

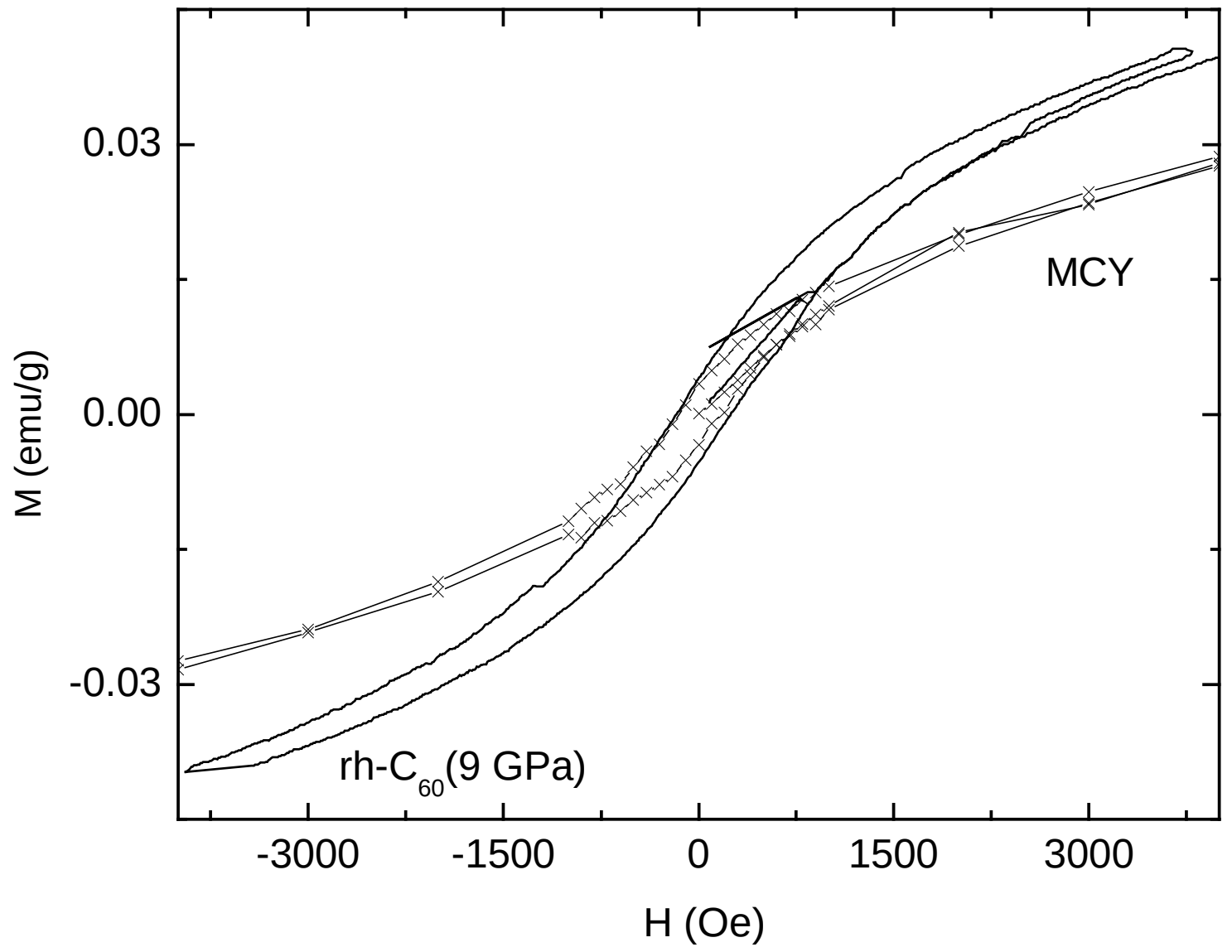

Figure 3. M(H) hysteresis loops obtained at T = 10 K for MCY and rh-$C_{60}$ samples [6].

In addition, Fig. 4 illustrates that the high field portion of the magnetization curves can be fitted by the equation $M(H) = M_s + \chi H$ with the same paramagnetic susceptibility $\chi \approx 3\cdot10^{-6}$ emu/g·Oe for both [MCY and rh-$C_{60}$ (9 GPa)] samples; $M_s$ is the spontaneous magnetization obtained from the extrapolation of the linear M(H) region to H = 0.

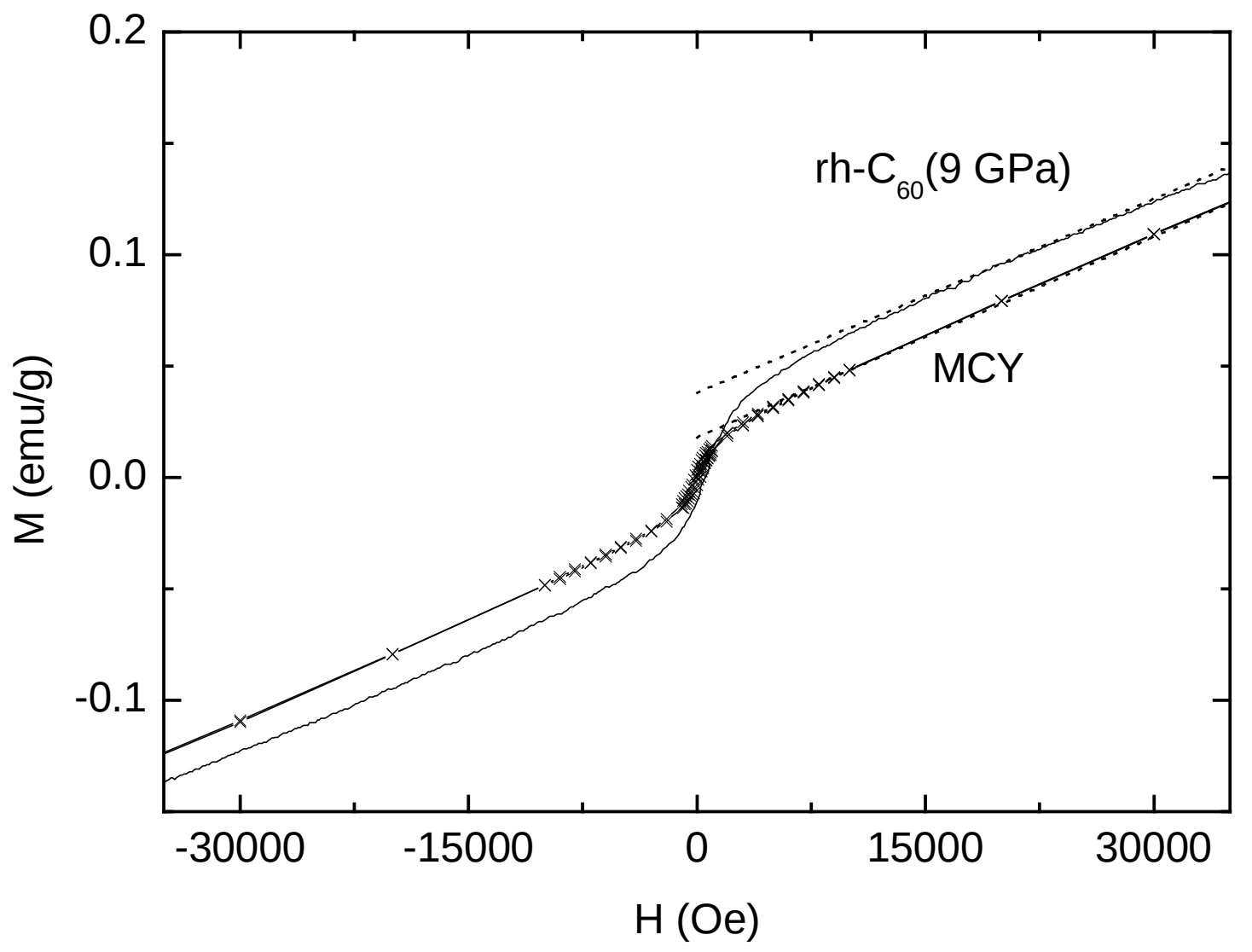

Figure 4. M(H) obtained at T = 10 K for MCY and rh-$C_{60}$ [6] samples; dotted lines are obtained from the equation $M(H) = M_s + \chi H$, where $M_s = 0.018$ emu/g (MCY), $M_s = 0.038$ emu/g (rh-$C_{60}$), and $\chi \approx 3\cdot10^{-6}$ emu/g·Oe for both rh-$C_{60}$ and MCY samples.

The above analysis strongly suggests that the ferromagnetism in microporous carbon and nominal rh-$C_{60}$ samples has a common origin, possibly associated with fullerene-like fragments. There exists both experimental and theoretical evidence that fullerene-like fragments with positive and/or negative curvature should be a common feature of microporous carbons, indeed [20]. Also, such fragments should naturally appear in rh-$C_{60}$ samples with partially destroyed $C_{60}$ molecules [21, 5-7]. On the other hand, since HOPG may have a small (if any) number of topological defects, the experimental observation of a much smaller ferromagnetic magnetization is not unexpected.

### 3. OXYGEN ADSORPTION EFFECT ON THE MAGNETIC PROPERTIES OF GRAPHITE

In 1996, Murakami and Suematsu [22] reported on the ferromagnetic ordering at T ≤ 800 K in $C_{60}$ crystals exposed to light in the presence of oxygen gas. Recent measurements [23] corroborated the occurrence of high temperature ferromagnetism in $C_{60}$ under photo-assisted oxidation. It has been suggested [22, 23] that the photo-polymerization plays an important role to the ferromagnetism occurrence.

Aiming to shed more light on the relevance of the oxygen presence to the ferromagnetism in carbon-based materials, we explored the effect of oxygen and other adsorbed gases on the magnetic properties of a graphite powder.

In these experiments, an activated graphite powder was prepared by cutting and grinding a graphite rod at T = 300 K in different atmospheres: Ar, He, $N_2$, $H_2$, $O_2$, and air. The graphite rod was from Carbon of America Ultra Carbon, sell by Alfa Aesar (stock No. 40766), AGKSP grade, (ultra "F") 99.9995 % purity or 5 ppm of total impurities and maximum of 1 ppm of impurities per element. The powder was produced by cutting and grinding the graphite rod on the edge and side area of a new and clean circular diamond saw blade. The cutting and grinding system was inside a plastic bag filled with the gas. Using a gas hose, a continuous gas stream was also forced to blow in the grinding area.

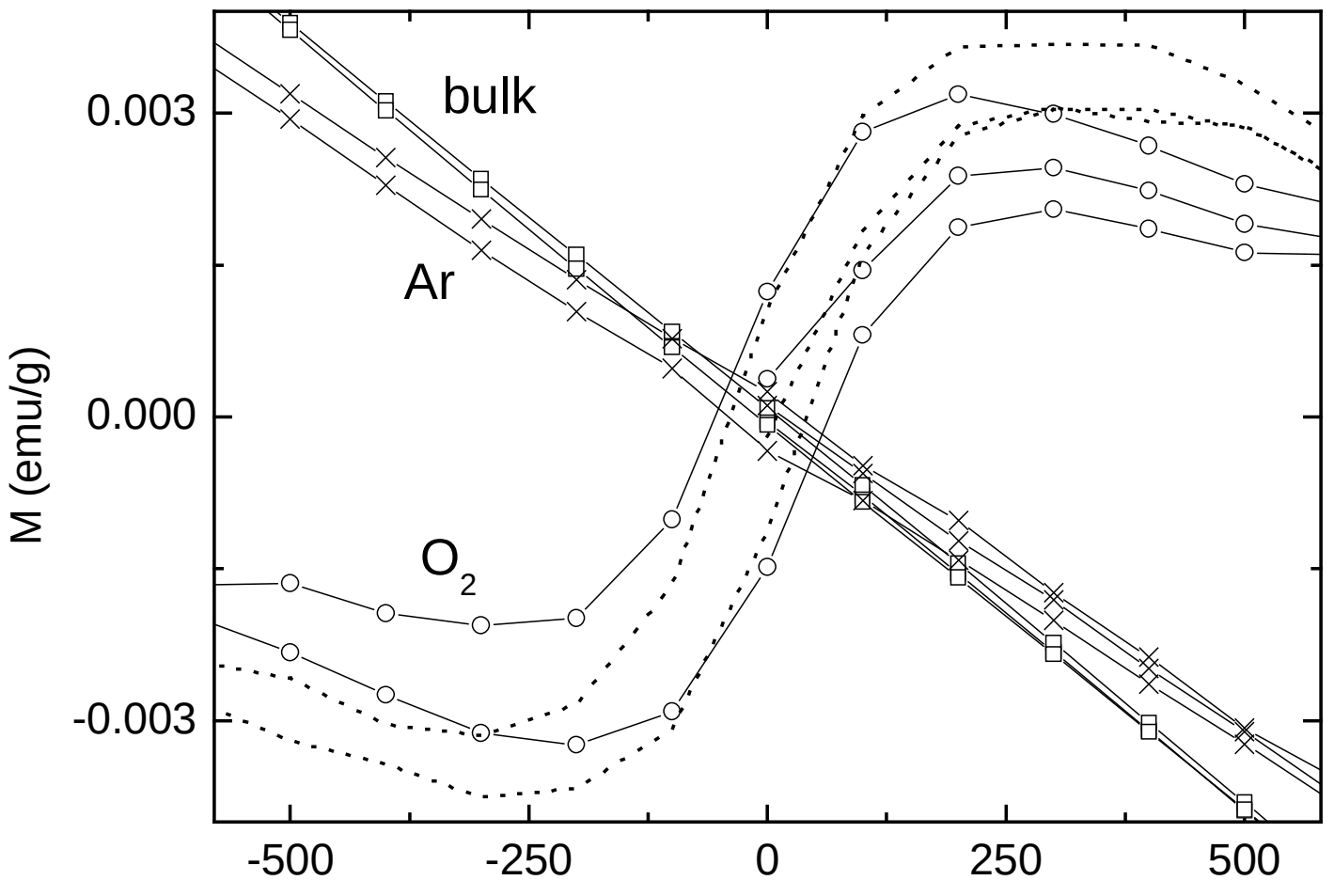

Figure 5. M(H) measured at T = 10 K for the graphite samples: bulk (virgin) sample and powdered samples prepared under argon (Ar) and oxygen ($O_2$) atmospheres; dotted line corresponds to M(H) measured for the oxidized sample at T = 300 K.

From the x-ray diffraction spectrum we estimate the graphite crystallites size $L_a \approx$ 1000 Å and $L_c \approx$ 400 Å. The powder particle size was found to be 150 ± 70 μm, using sieves.

Figure 5 illustrates that M(H) curves measured for the virgin bulk sample as well as the powdered under Ar atmosphere sample practically coincide and demonstrate the sample diamagnetism and a negligible magnetic hysteresis. Similar results were obtained for the graphite powders prepared under He, $N_2$, and $H_2$ gas environment.

At the same time, the samples prepared under oxygen exposure showed a pronounced ferromagnetic response. This can be seen in Fig. 5 where ferromagnetic hysteresis loops measured for the powder prepared under oxygen exposure are clear. This observation provides unambiguous evidence that the ferromagnetism can be triggered by the presence of oxygen. Figure 6 illustrates another important experimental fact, namely that the oxygen effect is reversible: the ferromagnetism vanishes with time after the sample is removed from the oxygen atmosphere.

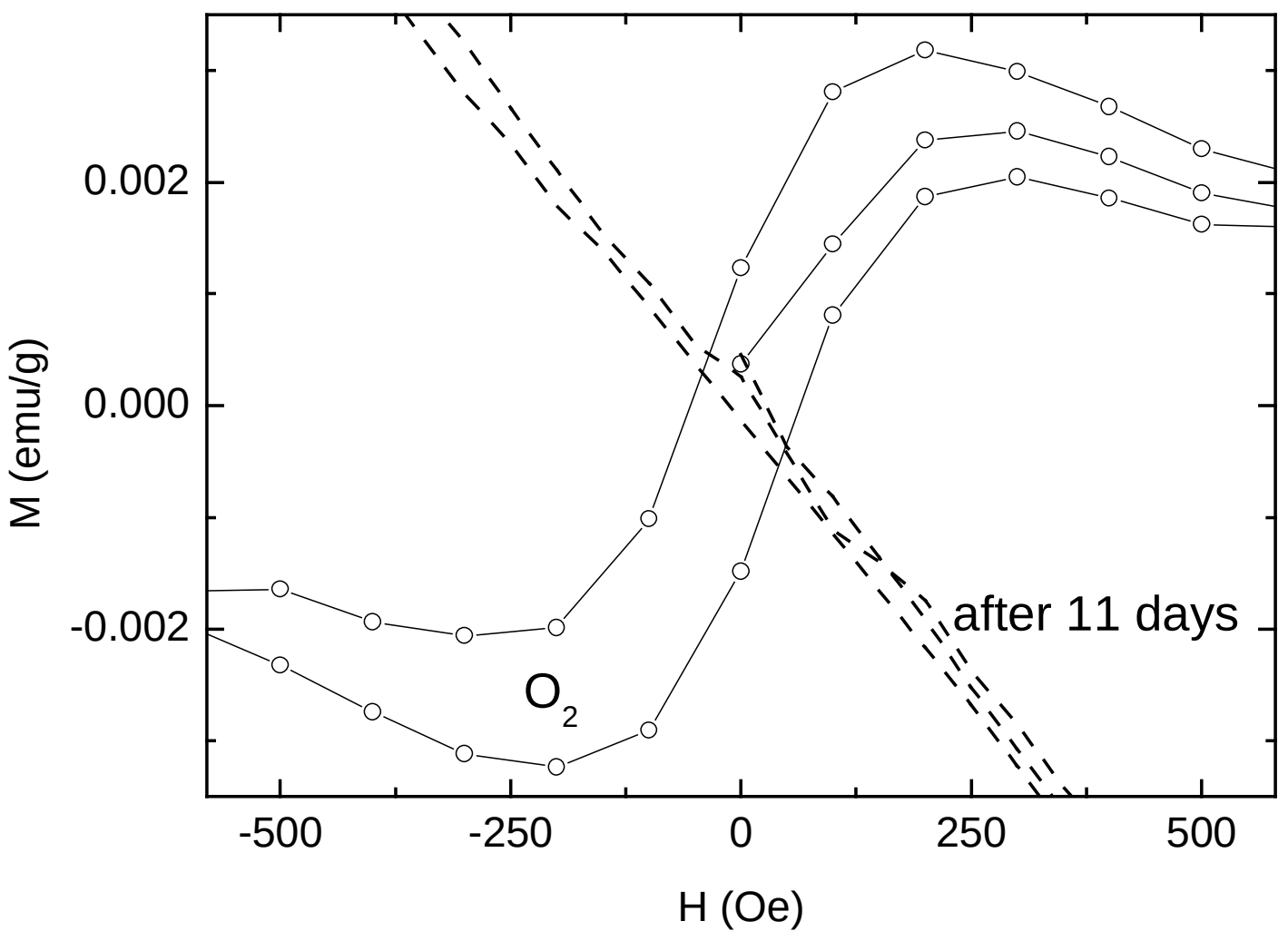

Figure 6. M(H) for the graphite powder prepared in oxygen ($O_2$) atmosphere measured at T = 10 K right after preparation and after 11 days of the sample staying in air.

Summarizing, the results given in Fig. 5 and Fig. 6 suggest that an adsorption followed by desorption of oxygen gas at the graphite surface are responsible for the appearance and vanishing of the ferromagnetism, respectively. Note that the process used here to prepare the graphite powder samples is expected also to introduce a large number of structural defects.

Previous studies on the oxidation of graphite, carbon nanotubes (graphite sheets wrapped into cylinders) [24-26] and related carbon materials revealed, in particular, that the presence of defects at the sample surface is needed for the oxygen adsorption.

Taking all the experimental evidence together, one may speculate that the ferromagnetism enhancement observed, e. g., in strongly disordered microporous carbon, see section 2, is related to oxygen trapped at defect sites. On the other hand, our previous work [2] revealed that a low-vacuum heat treatment of HOPG samples can either enhance the ferromagnetic response or trigger superconducting-like M(H) hysteresis loops even at room temperature, once again suggesting that adsorbed gases play a crucial role in the anomalous magnetic behavior of graphite.

Figure 7 (a, b) presents our recent data obtained at room temperature as before [2] but now for a different HOPG sample providing additional evidence for the interplay between ferromagnetic and superconducting behavior of graphite subjected to a heat treatment. As Fig. 7 (a) illustrates, the ferromagnetic-type (FM) hysteresis loop M(H) measured immediately after the heat treatment [the HOPG sample was kept at T = 870 K under low vacuum (~ 0.05 mbar) during 2 hours] has been transformed to a loop characteristic of type-II superconductors (SC) when H || c. The change in the magnetic response has occurred after 14 days of keeping the sample in air at ambient conditions.

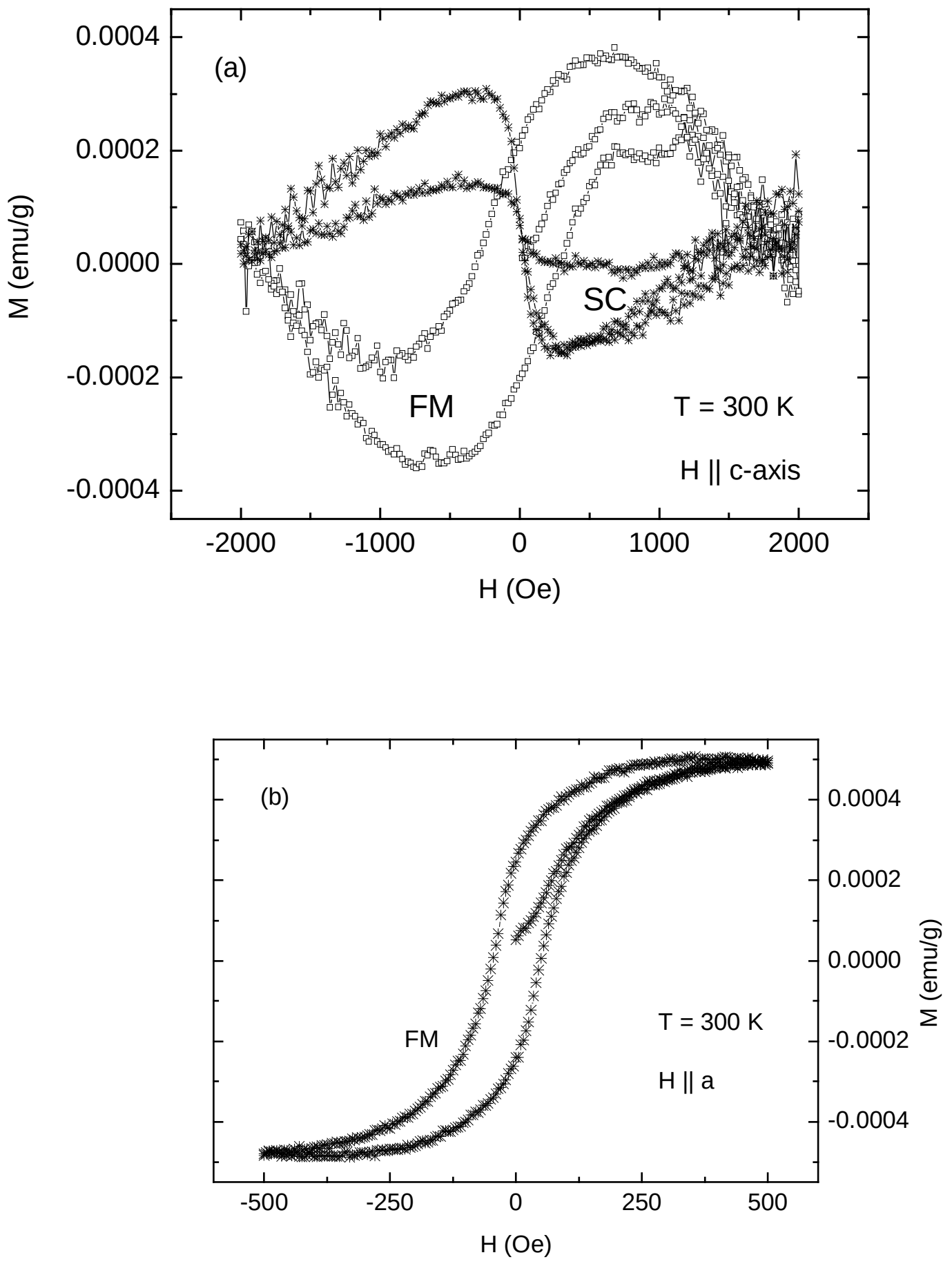

Figure 7. (a) M(H) measured for a HOPG sample from Union Carbide Co. immediately after the heat treatment (FM), and after 14 days keeping the sample at ambient conditions (SC) when H || c. The curves were obtained subtracting the diamagnetic background magnetization $M_0(H) = -\chi H$ ($\chi = 1.61 \cdot 10^{-5}$ emu/g·Oe) from the measured signal; (b) M(H) hysteresis loop obtained with H || graphite planes (H || a) for the SC sample after the background signal subtraction ($\chi = 2.3 \cdot 10^{-6}$ emu/g·Oe).

Noting, the FM → SC transformation recorded in H || c measurements, Fig. 7 (a), did not affect the FM response measured with the field applied parallel to the graphite planes (H || a), see Fig. 7 (b), suggesting that the superconductivity is localized within the sample basal planes. The results provide also evidence for the coexistence of superconducting and ferromagnetic states.

Also, we have studied the stability and sensitivity of this (transformed) superconducting state to additional sample heating. Figure 8 shows the superconducting-type hysteresis loop M(H) of Fig. 7 (a) together with the hysteresis loop measured immediately after heating the sample up to 350 K (in situ) (H || c). As can be seen from Fig. 8, this slight heating already results in a reduction of the hysteresis loop width.

We speculate that both the aging effect, i. e. the time dependence of the sample magnetic response, and its sensitivity to relatively small temperature variation may be related to gasses (most likely oxygen) adsorption-desorption and/or its migration to different defect sites.

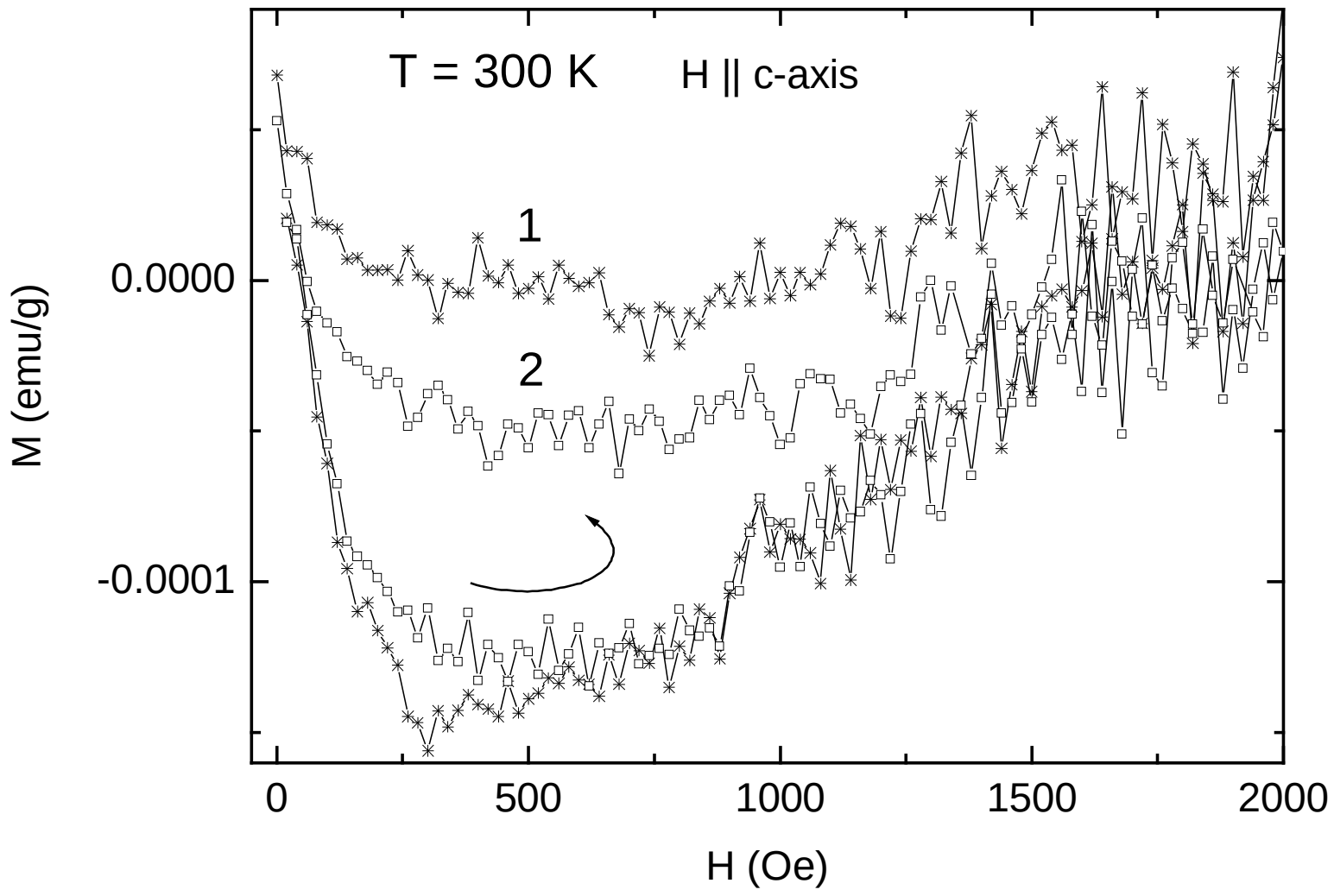

Figure 8. Superconducting-like M(H) hysteresis loops: curve 1 is the same as in Fig. 7, and curve 2 is measured after heating the sample up to 350 K. The curves were obtained subtracting the diamagnetic background magnetization $M_o(H) = -\chi H$ ($\chi = 1.61\cdot 10^{-5}$ emu/g·Oe) from the measured signal.

Furthermore, our studies revealed that similar phenomena, i. e. the interplay between ferromagnetic and superconducting order parameters and aging effects take also place in sulfur-doped graphite samples. Results of these studies are given in the next section.

## 4. COEXISTENCE OF SUPERCONDUCTING AND FERROMAGNETIC INSTABILITIES IN GRAPHITE-SULFUR COMPOSITES.

Graphite-sulfur (C-S) composites were prepared [11] by mixing graphite powder consisting of ~ 8 μm size particles [the impurity content in ppm: Fe (32), Mo (< 1), Cr (1.1), Cu (1.5)] and sulfur powder (99.998 %; Aldrich Chemical Company, Inc.) in a ratio C:S = 1:1. The mixture was pressed into pellets, held under Ar atmosphere at 650K for one hour and subsequently annealed at 400 K for 10 hours before cooling to room temperature. The final sulfur contents in the composite was 23 wt %. X-ray (θ - 2θ geometry) measurements revealed a small decrease in the c-axis parameter of the hexagonal graphite from c = 6.721 Å in the pristine graphite powder to c = 6.709 Å in the composite sample, and no changes in the lattice parameters of the orthorhombic sulfur (a = 10.45 Å, b = 12.84 Å, c = 24.46 Å). No impurity or additional phases were found [11].

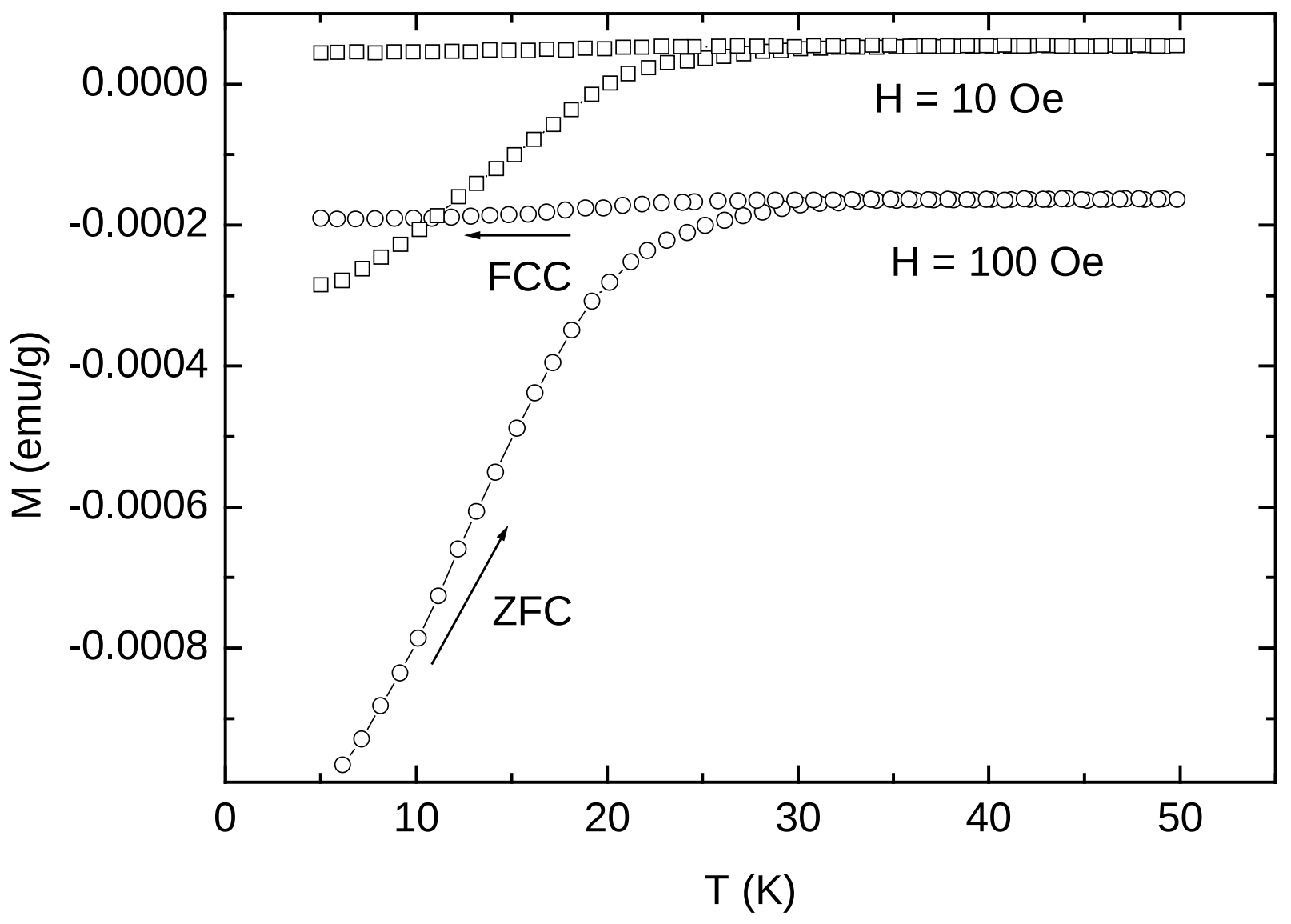

Figure 9. Temperature dependencies of magnetization, M(T), measured in graphite-sulfur composite (CS1) in zero-field-cooled (ZFC) and field-cooled on cooling (FCC) regimes at two applied fields; 10 Oe and 100 Oe.

Figure 9 presents magnetization M(T) measured in as-received sample (labeled here as CS1) for applied fields H = 10 Oe and H = 100 Oe. The magnetization data corresponding to the zero-field-cooled (ZFC) regime, $M_{ZFC}(T)$, were taken on heating after the sample cooling at zero applied field, and the magnetization in the field-cooled on cooling (FCC) regime, $M_{FCC}(T)$, was measured as a function of decreasing temperature in the applied field. Figure 10 gives a detailed view of the data obtained for H = 100 Oe in the vicinity of $T_c$(H = 100 Oe) = 33 K below which a departure of $M_{ZFC}(T)$ from $M_{FCC}(T)$ takes place. As can be seen from this plot, both $M_{ZFC}(T)$ and $M_{FCC}(T)$ become more diamagnetic at $T < T_c(H)$.

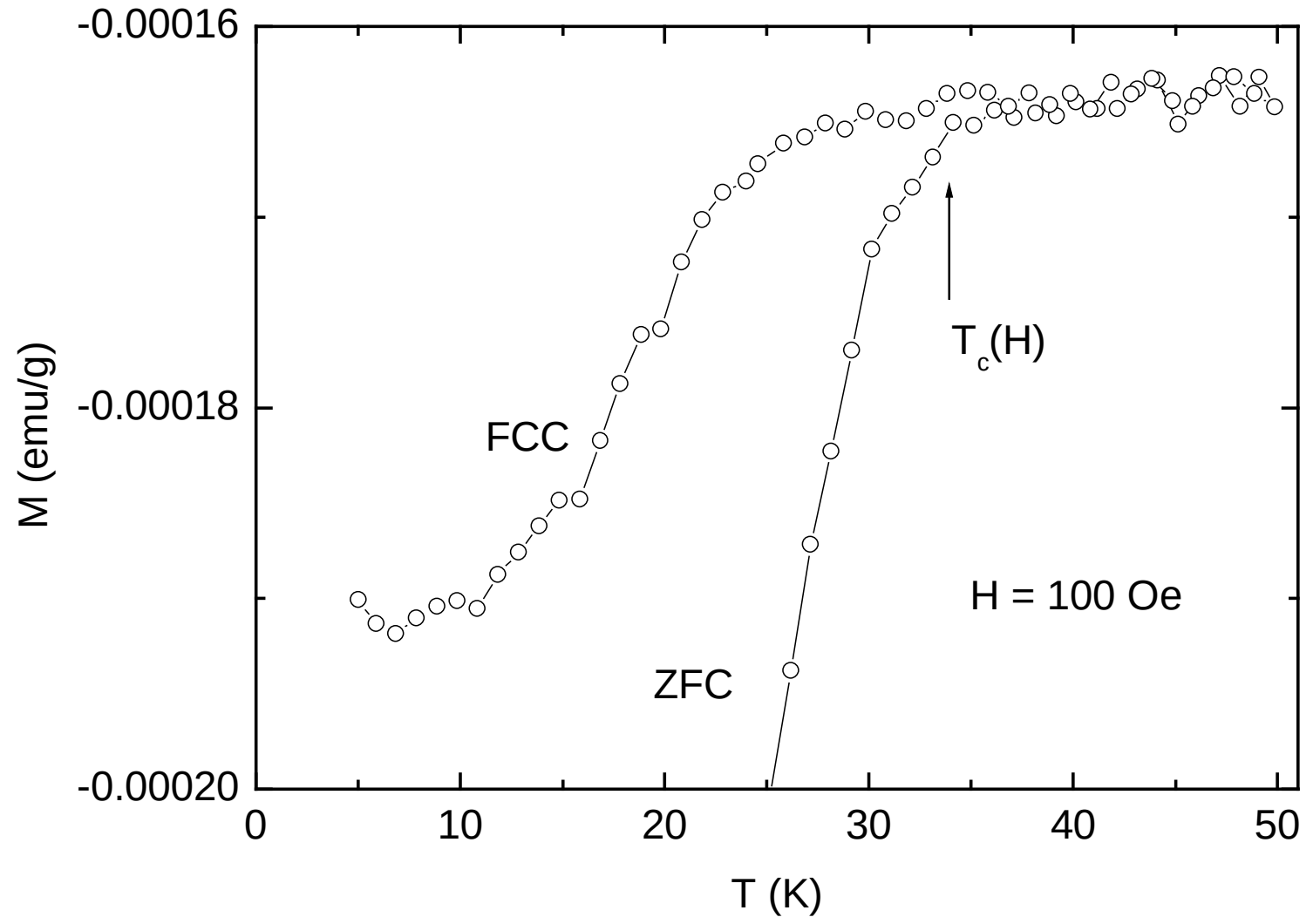

Figure 10. Enlarged view of the superconducting transition recorded for the CS1 sample at H = 100 Oe.

Such magnetization behavior is characteristic of superconductors: enhancement of the diamagnetism below the superconducting transition temperature $T_c(H)$ originates from the screening supercurrents (ZFC regime) and the Meissner-Ochsenfeld effect of magnetic flux expulsion (FFC regime). The magnetization data of Fig. 9 and Fig. 10 demonstrate that the sample superconducting volume fraction is rather small. We estimate it as ~ 0.05 % of the value expected for a bulk ideal superconductor [11].

It can also be seen in Fig. 9 that as the applied field increases, the normal state orbital diamagnetism of graphite overcomes the paramagnetic (ferromagnetic) contribution to the magnetization resulting in a negative total magnetization above $T_c$.

As expected for superconductors, the diamagnetic magnetization is suppressed by the applied magnetic field, see Fig. 11.

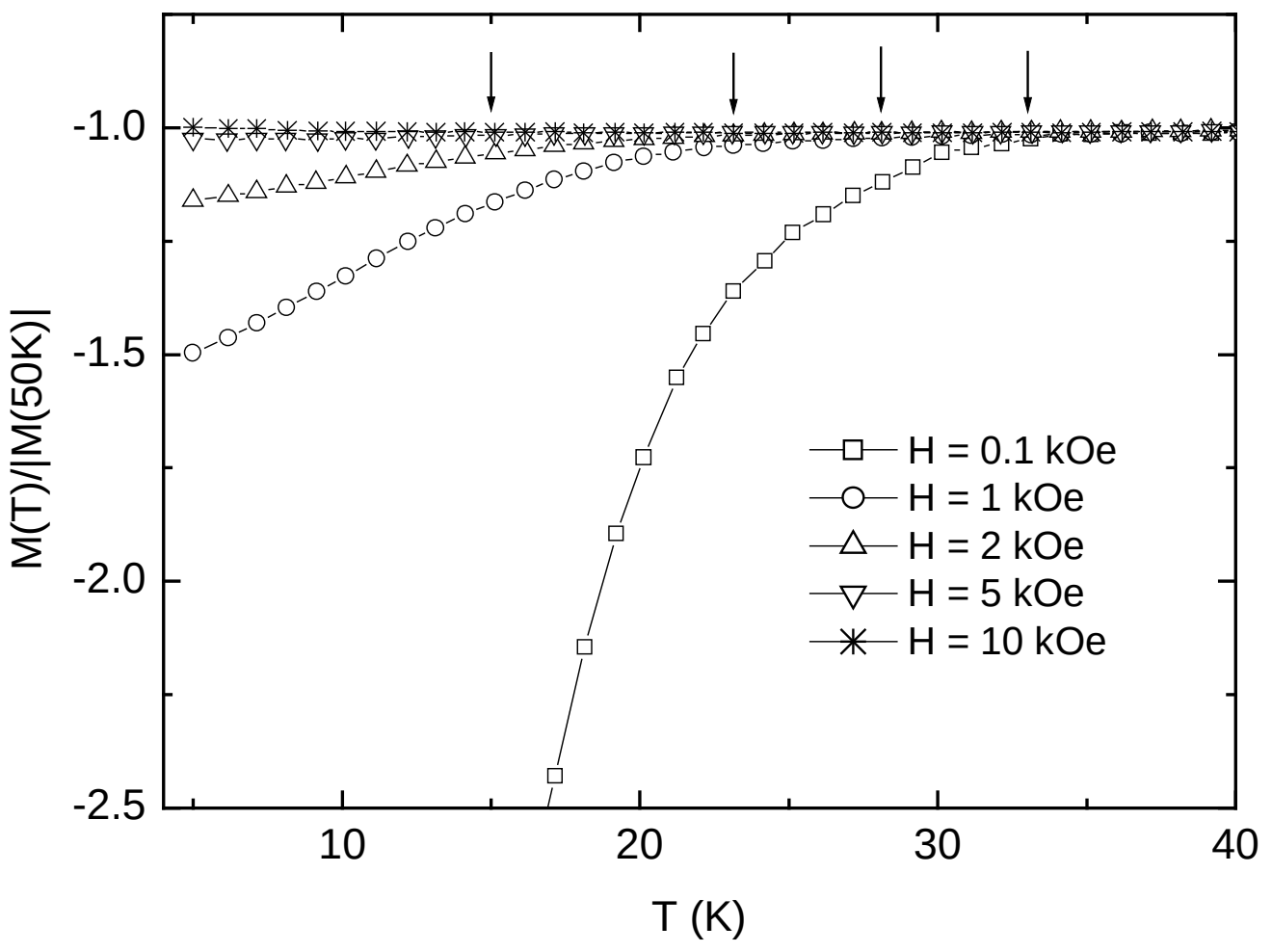

Figure 11. Normalized ZFC magnetization measured in the sample CS1 at various applied fields. Arrows denote the superconducting transition temperature $T_c$(H).

Figure 12 presents M(H) measured for the CS1 sample at T = 6 K. The occurrence of magnetic hysteresis characteristic of type-II superconductors can be seen for this sample even without subtraction of the orbital diamagnetic contribution.

Similar to experiments with oxidized graphite, see section 3, the magnetic response of C-S composites is found to be time-dependent. To illustrate this we plotted in Fig. 13 the reduced magnetization measured for the same C-S sample with two weeks time interval. As can be seen from Fig. 13, the superconducting response has diminished considerably after keeping the sample at ambient conditions.

Dozens of C-S samples with different $T_c$ and superconducting volume fraction were obtained. The results proved that powdering of graphite before the sulfur reaction is of importance for the superconductivity occurrence. Both superconductivity and ferromagnetism were observed in the C-S samples. Interestingly, the character (FM or SC) as well as the magnitude of the magnetic response can change (decrease or increase) with time continuously or abruptly [13].

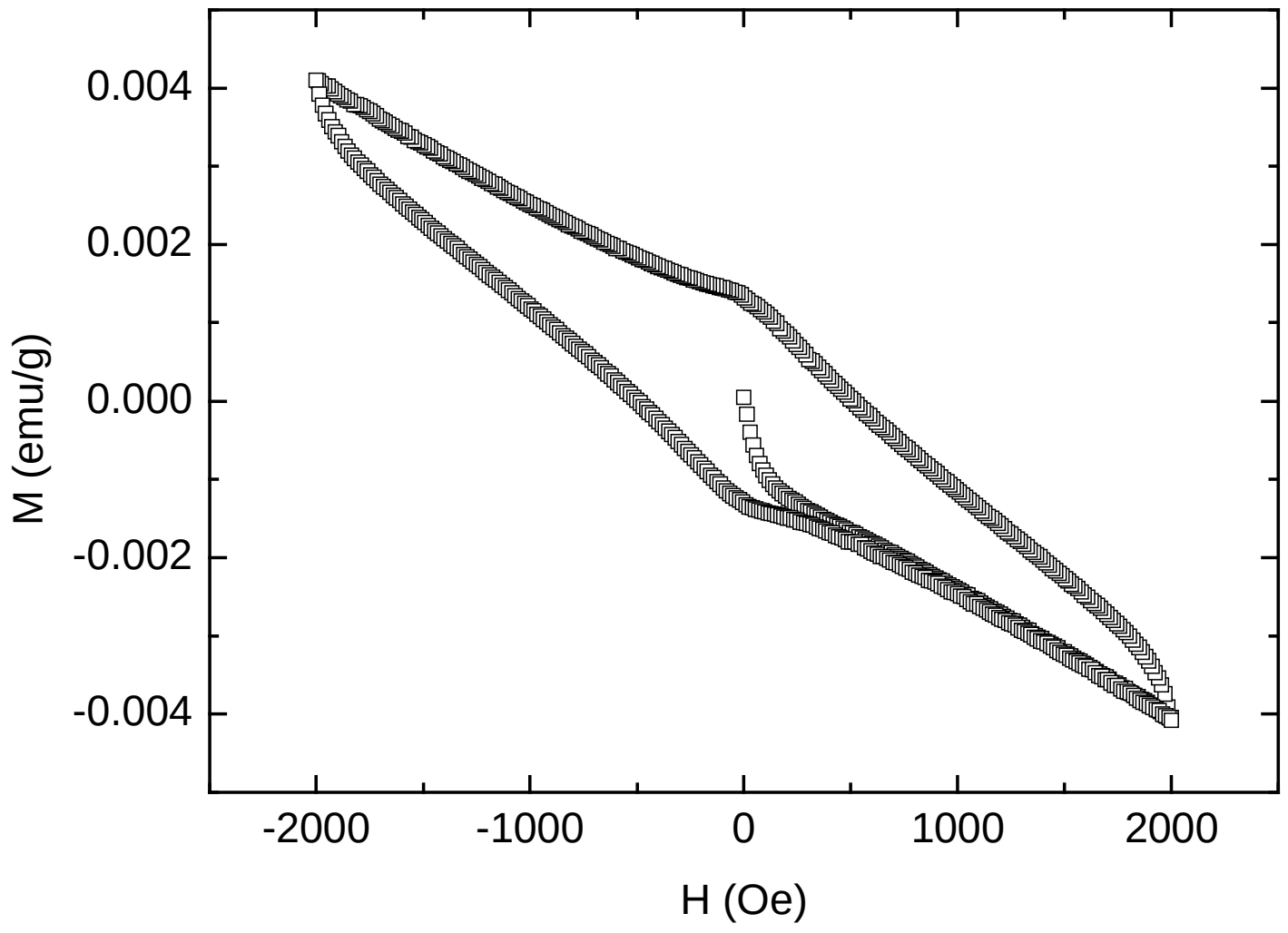

Figure 12. Magnetization hysteresis loop M(H) measured for CS1 sample at T = 6 K.

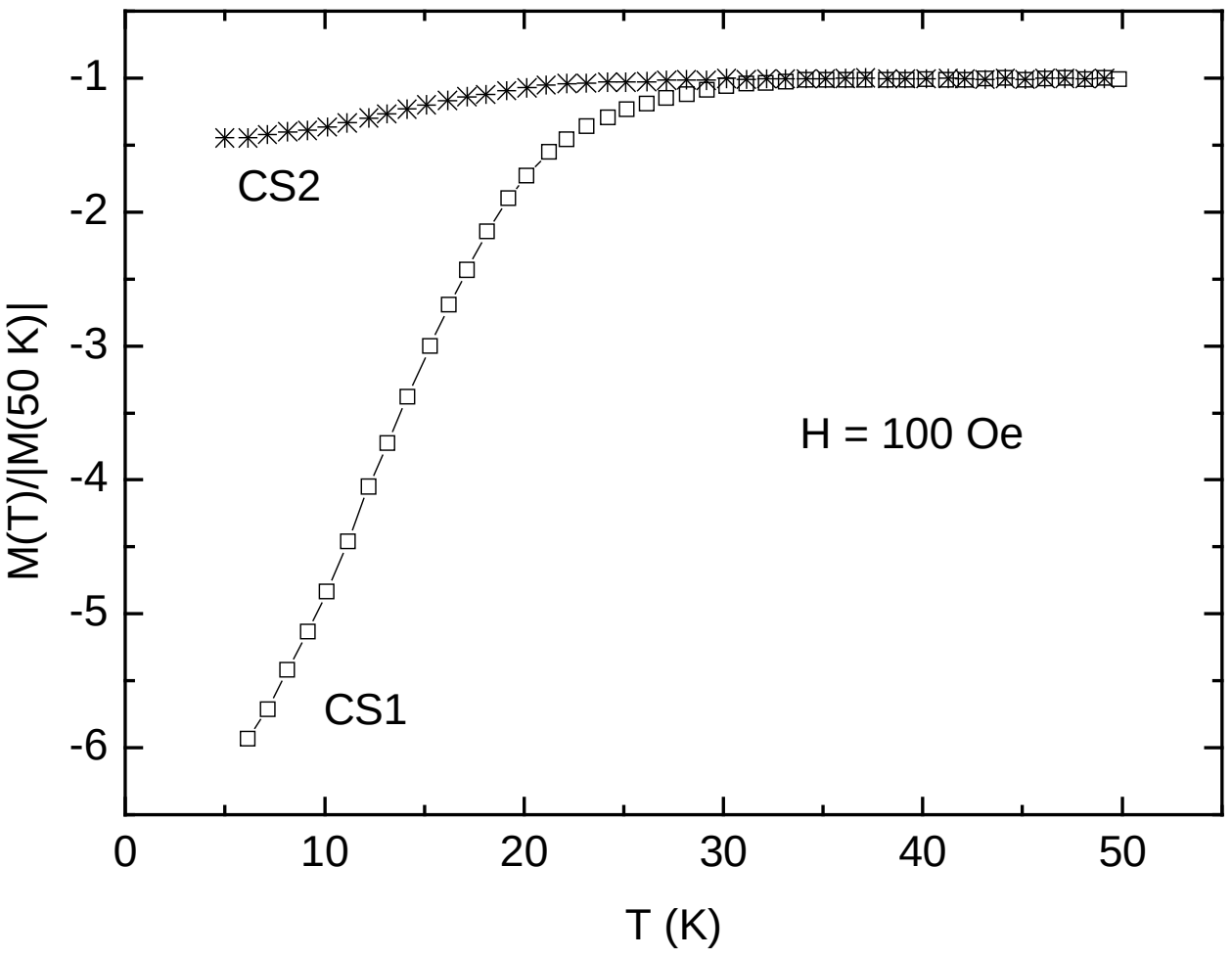

Figure 13. Normalized ZFC magnetization measured for the same graphite-sulfur composite sample immediately after its synthesis (CS1) and after 14 days keeping the sample at ambient conditions (CS2).

In what follows we present detailed experimental evidence for the coexistence of superconductivity and ferromagnetism in graphite-sulfur composites. The data were obtained on a C-S sample (labeled here as CS3) with $T_c(0) = 9$ K whose ferromagnetic properties were unchanged during 5.5 months and remains superconducting until today (more than 3 years after preparation).

The CS3 sample was prepared using graphite rods from Carbon of America Ultra Carbon, AGKSP grade, ultra "F" purity (99.9995%) (Alfa-Aesar, # 40766) and sulfur chunks from American Smelting and Refining Co. that are spectrographically pure (99.999+ %). A pressed pellet ($\phi$ = 6 mm, ~7000 lbf) of graphite was prepared by pressing graphite powder that was produced by cutting and grinding the graphite rod on the edge and side area of a new and clean circular diamond saw blade. The graphite pellet was encapsulated with sulfur chunks (mass ratio ~ 1:1) in quartz tube under 1/2 atmosphere of argon and heat treated in a tube furnace at 400 °C for one hour and then slowly cooled (4 °C/h) to room temperature. X-ray diffraction measurements ($\theta$-2$\theta$ geometry and rocking curves) of the reacted sample yielded a spectrum with only the superposition of the (00ℓ) diffraction peaks of graphite with the orthorhombic peaks of sulfur with no extra peak due to a compound, second phase or impurity. The c-axis lattice parameter (c = 6.72 Å) of the sample is equal to the pristine graphite powder pellet, which precludes sulfur intercalation. The diffraction pattern also shows a (00ℓ) preferred orientation, which was confirmed by rocking curve scans that yield a $\Delta\theta = 6°$ (FWHM) for the (002) peak, due to the highly anisotropic (plate-like) shape of the graphite grains. The sample (~ 5 x 2.5 x 1.7 $mm^3$) was cut from the reacted pellet and used for the magnetic moment measurements as well as the above described analyses. A lower limit for the superconducting volume fraction was estimated as 0.02 % [13].

Shown in Fig.14(a) are temperature dependencies of the ZFC magnetization M(T) after subtraction the normal state value at T = 10 K, M(10K), measured for various magnetic fields applied perpendicular to the largest surface of the sample (H || c). It can be seen from Fig. 14(a) that the difference |M(T)-M(10K)| increases below the superconducting transition temperature $T_c = 9$ K. At the same time, M(T) measurements with the applied magnetic field parallel to the main sample surface (H || a) yield a different magnetic response. As Fig. 14(b) illustrates, no sign of the superconducting transition could be detected within the data noise of ~ 5 x $10^{-6}$ emu/g. Note that the scale ranges in Fig. 14(b) and Fig. 14(a) are practically the same. These results indicate that the superconducting state is highly anisotropic and is associated with the graphite planes.

Figure 15 shows magnetization hysteresis loops $M(H)-M_o(H)$ measured with the ZFC procedure and H || c for T = 7, 8, 9 and 10 K, where $M_o = \chi H$ ($\chi = -7.12 \times 10^{-6}$ emu $g^{-1}$ $Oe^{-1}$) is the diamagnetic background signal. Figure 15(a) demonstrates a characteristic of type-II superconductors hysteresis loop obtained at T = 7 K. As temperature increases above $T_c = 9$ K, the measured hysteresis loops resemble those of ferromagnetic materials, see Fig. 15 (d). For temperatures at and just below $T_c$, Fig. 15 (b, c), the presence of both superconducting and ferromagnetic contributions to the measured magnetization can be seen.

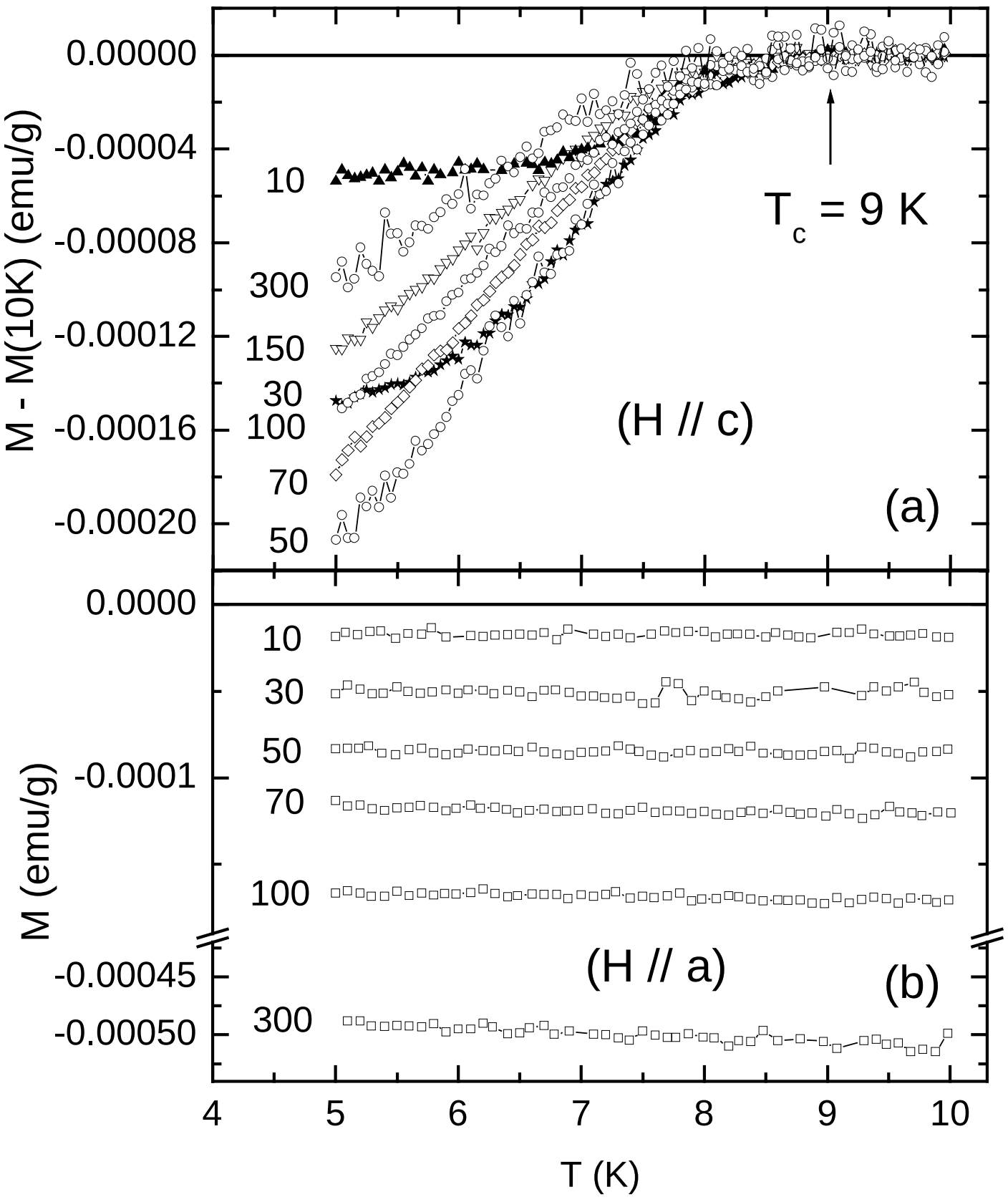

Figure 14. (a) temperature dependencies of ZFC magnetization M(T) - M(T =10K), obtained for various magnetic fields H ‖ c: (▲), H = 10 Oe; (★), H = 30 Oe; (○), H = 50 Oe; (◊), H = 70 Oe; (○), H = 100 Oe; (∇), H = 150 Oe; (○), H = 300 Oe; (b) temperature dependencies of ZFC magnetization obtained with H ‖ a for various magnetic fields, as indicated in Oe next to each curve.

Thus, results presented in Fig. 15 provide evidence for the coexistence of superconductivity and ferromagnetism in the C-S.

Again, for the M(H) measurements with the applied magnetic field parallel to the graphite planes (H ‖ a), a different magnetic response is obtained. This can be seen in Figure 16 (a-c) which presents M(H) obtained for T = 5, 7 and 9 K after subtraction of the linear diamagnetic background signal $M_o = \chi H$, where $\chi = -2.25 \times 10^{-6}$ emu $g^{-1}$ $Oe^{-1}$: three almost identical ferromagnetic-type hysteresis loops are clearly seen both in the superconducting and normal states (the ferromagnetic behavior of C-S composites persists well above the room temperature [13]).

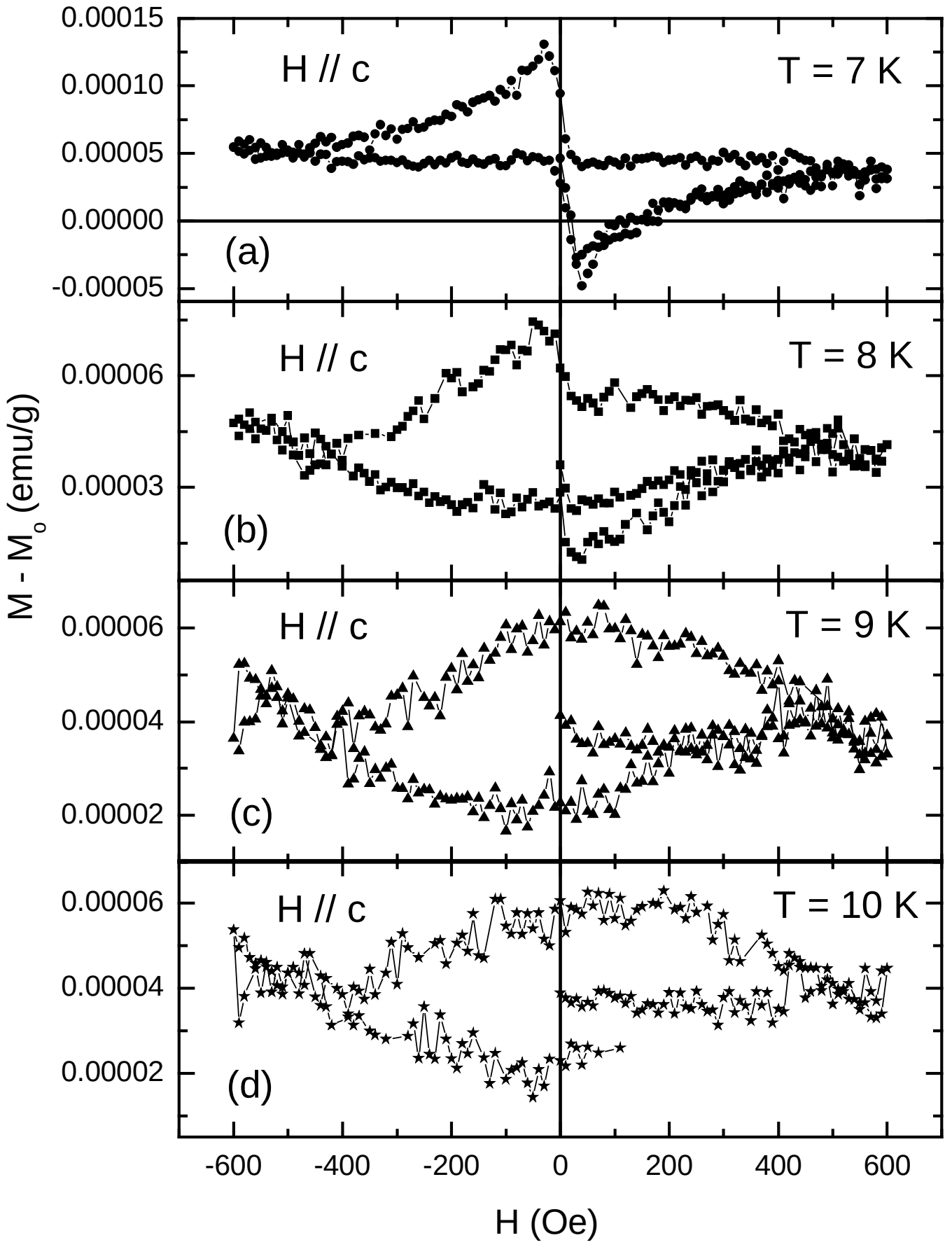

Figure 15. Magnetization hysteresis loops M(H) measured for CS3 sample with H || c after subtraction of the diamagnetic background signal $M_o = \chi H$ ($\chi = -7.12 \times 10^{-6}$ emu $g^{-1}$ $Oe^{-1}$) at (a) T = 7 K, (b) T = 8 K, (c) T = 9 K and (d) T = 10 K.

Noting also that angle dependent magnetization measurements performed on the CS3 sample [14] provided evidence for the interaction between superconducting and ferromagnetic order parameters [14]. The interaction revels itself through a rotation of the ferromagnetic moment direction by 90$^\circ$ at $T < T_c(0)$. This fact is particularly intriguing taking into account the smallness of the measured total superconducting signal.

It is interesting to compare $T_c(H)$ measured for CS1 and CS3 samples. The data given in Fig. 17 illustrate that $H(T_c)$ in the vicinity of the zero-field transition temperature $T_c(0)$ can be very well described by the power law $H = H_0[1 - T_c(H)/T_c(0)]^{3/2}$ for both samples.

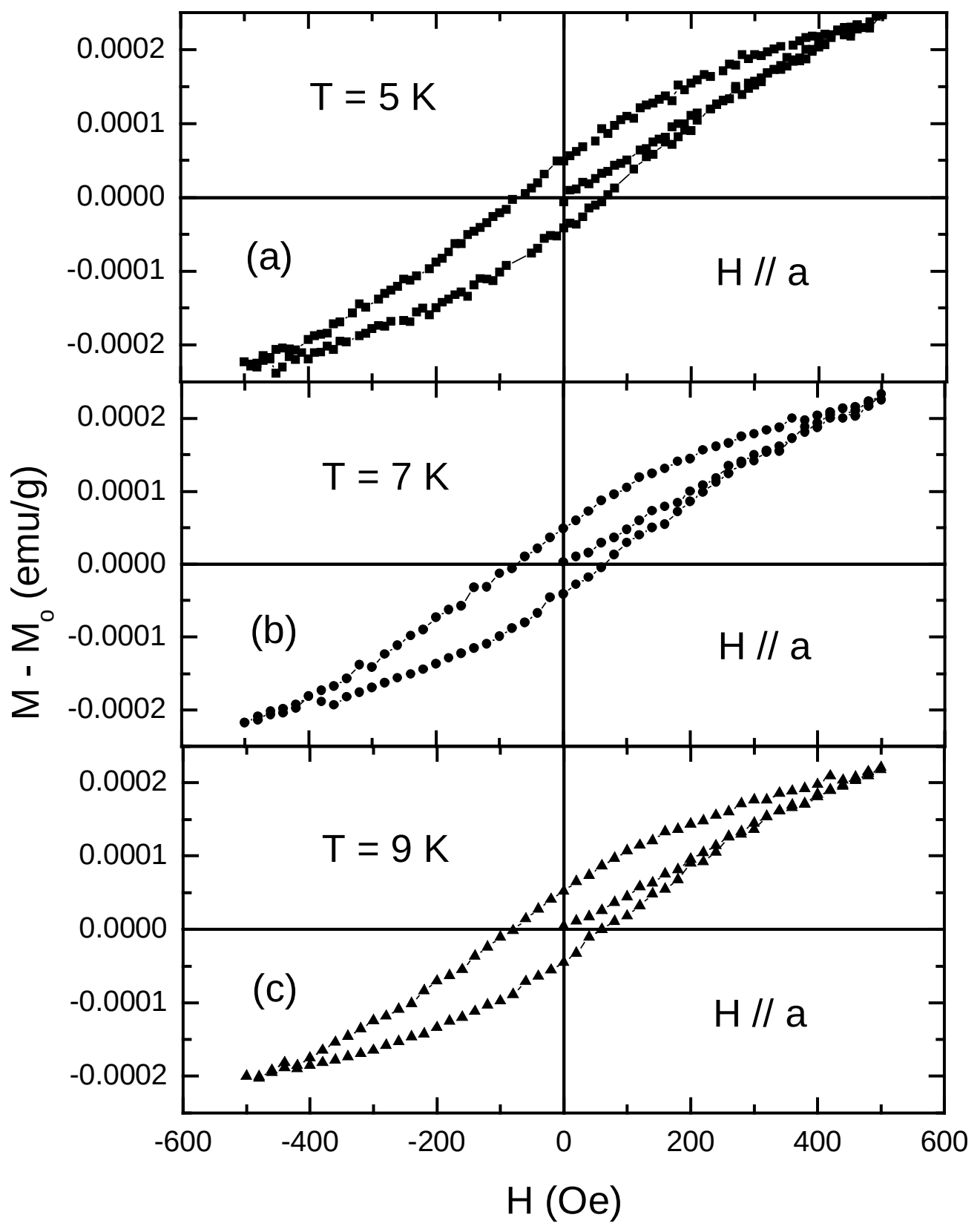

Figure 16. “In-plane” (H || a) hysteresis loops M(H) – $M_0$(H), $M_o = \chi H$, $\chi = -2.25 \times 10^{-6}$ emu $g^{-1}$ $Oe^{-1}$, obtained for CS3 sample at (a) T = 5 K, (b) T = 7 K and (c) T = 9 K.

This suggests a common origin for the superconducting transition in these samples in spite of their relatively large difference in $T_c(0)$. The observed nonlinear $H \sim (1 - t)^{3/2}$ dependence near $T_c(0)$ may be related to the upper critical field boundary associated with the Bose-Einstein condensation of preformed Cooper pairs [27]. Alternatively, the measured $H(T_c)$ boundary can be accounted for by the coupling between preexisting superconducting clusters that have a higher superconducting transition temperature [11, 13, 28]. Supporting this last scenario, $T_c(0)$ measured immediately after the CS3 sample synthesis was about 65 K which has declined with time and stabilized at $T_c(0)$ = 9 K after 11 days, see Fig. 18. Finally, we note that superconductivity in C-S samples is localized within de-coupled, on a macroscopic scale, clusters preventing the zero-resistance state of the macroscopic sample [11, 12, 13].

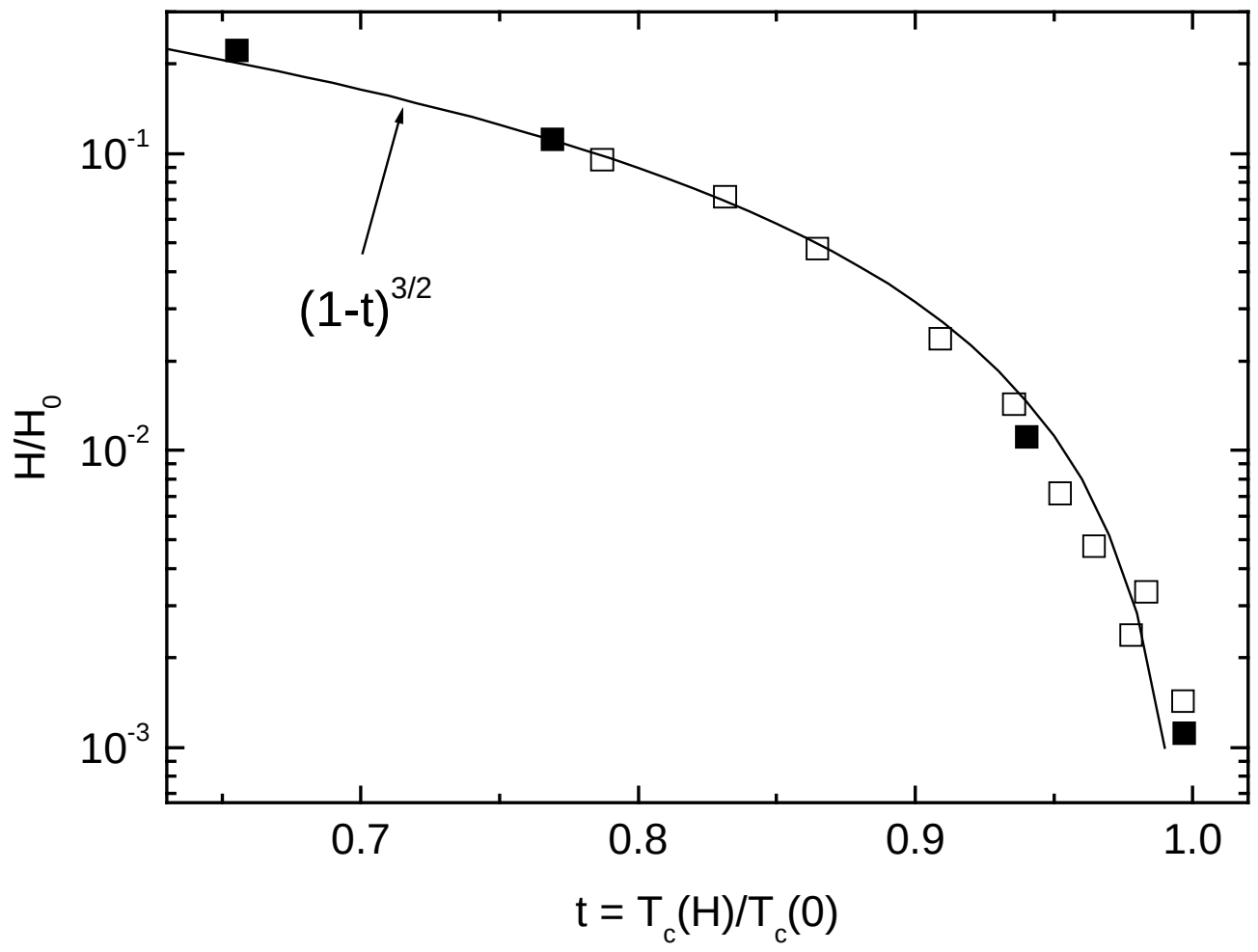

Figure 17. Reduced magnetic field $H/H_0$ vs. reduced temperature $t = T_c(H)/T_c(0)$ measured for CS1 (■) [11] and CS3 ( ) samples [13]; solid line corresponds to the equation $H/H_0 = [1-T_c(H)/T_c(0)]^{3/2}$ with $H_0$ = 9 kOe, $T_c(0)$ = 35 K for CS1, and $H_0$ = 21 kOe, $T_c(0)$ = 9 K for CS3.

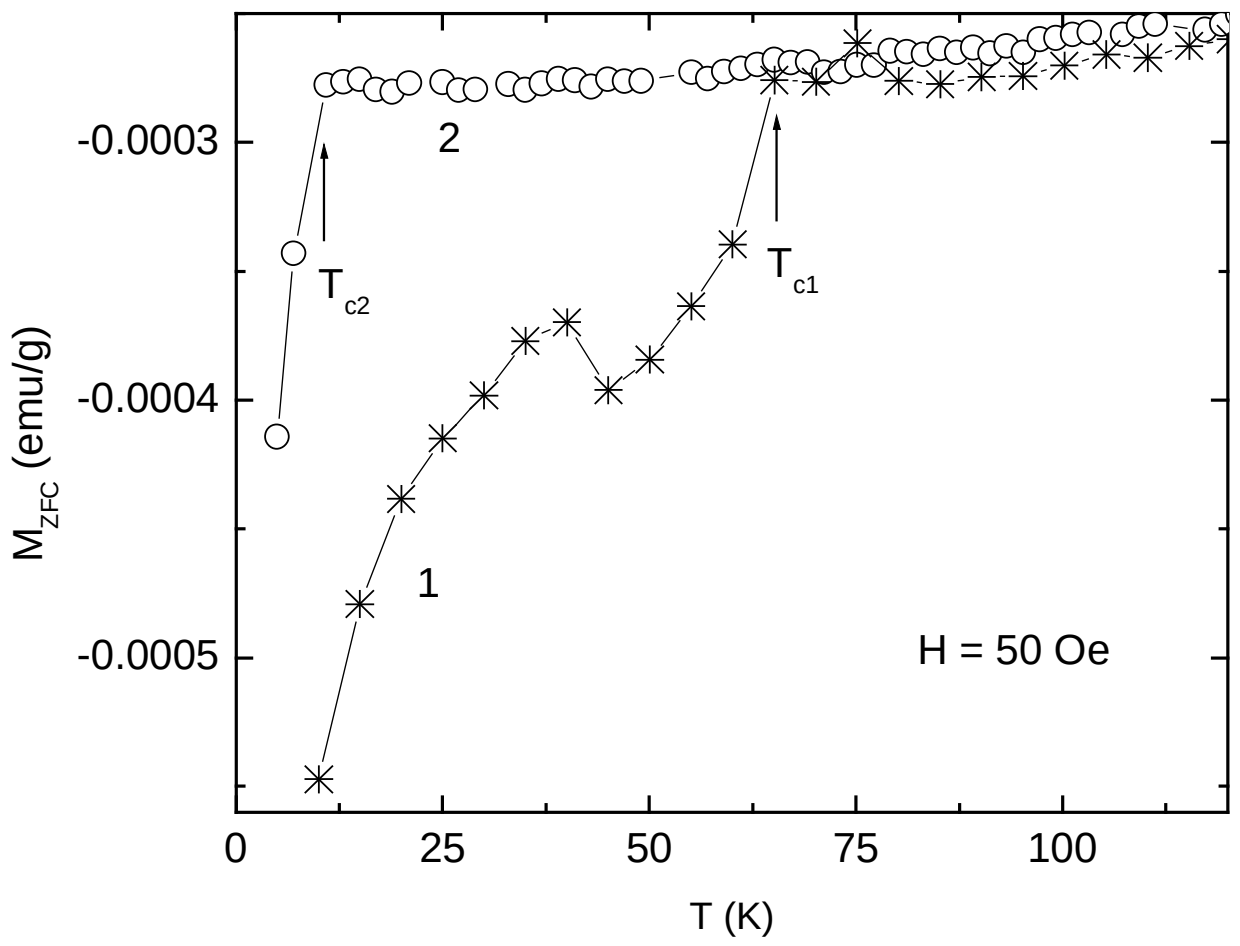

Figure 18. ZFC magnetization, $M_{ZFC}(T)$ measured at applied field H = 50 Oe for the CS3 sample immediately after the sample synthesis (curve 1) and after 11 days (curve 2).

## 5. DISCUSSION AND CONCLUDING REMARKS.

Because of the relatively weak ferromagnetic response of carbon-based materials, the possible effect of magnetic impurities, such as Fe, Ni, etc., cannot be ruled out a priori. This issue has been addressed by various researchers, see e. g. [29], who concluded that the measured carbon ferromagnetism cannot be accounted for by impurities. Such conclusion, however, does not exclude the possibility that magnetic impurities trigger ferromagnetism in the carbon system [8, 30, 31]. At the same time, the observations of ferromagnetism induced by graphite oxygenation, reported in this chapter, as well as the ferromagnetism triggered by proton irradiation [15] indicate that impurity-induced magnetism is not the whole story, at least. It is also clear, that the superconductivity in graphite caused by sulfur doping or re-arrangement of adsorbed oxygen on graphite surface has nothing to do with ferromagnetic impurities.

In order to clarify the nature of the observed phenomena, it is instructive to look more closely to previous studies of oxygenation of graphite and related carbon materials, see e. g. Refs. [24-26]. In particular, the results revealed an extreme sensitivity of the electrical resistance and thermoelectric power of carbon nanotubes (graphite sheets wrapped into cylinders) to oxygen (air) exposure [25]: the oxygen adsorption (1) decreases the nanotube resistance, and (2) it changes the thermoelectric power sign from negative to a positive one, suggesting that the oxygen-induced nanotube doping is with hole carriers [25, 26]. It has been demonstrated that the oxidation is triggered by the presence of graphite surface defects [25].

Adsorption of hydrogen atoms on graphite [32] also leads to a local charge enhancement as well as changes in the electronic structure. It has been theoretically shown [33] that the adsorption of atomic hydrogen opens a gap in the electronic spectrum of graphene in which a spin-polarized gap state is situated.

Furthermore, x-ray absorption spectra measured in C-S composites revealed that the outermost s-states of adsorbed S atoms interact with the graphite interlayer states [34].

All this makes us believe that a combined effect of structural disorder and adsorbed foreign atoms (molecules) such as S, H, O ($O_2$) can be behind the anomalous magnetic behavior of graphite and related carbon materials. Then, it is not unreasonable to assume that aging effects, including the FM $\rightarrow$ SC transformation, are related to migration of foreign elements on the sample surface. The very small (~ 0.01...0.05 %) Meissner as well as shielding effects can also be understood assuming the formation of superconducting patches at the graphite surface.

The current state of experiment does not allow us to discriminate unambiguously between theoretical models proposed to account for both high temperature ferromagnetism and superconductivity in graphite-based materials; for review articles see Refs. [3, 35, 36].

Nevertheless, we would like to recall here the mechanism for surface high temperature superconductivity proposed by Ginzburg [37]. In that model, conducting electrons (holes) interact with adsorbed atoms at the sample surface which leads to an effective attraction between the surface carriers. According to the BCS theory, $k_B T_c = \hbar\omega \exp[-1/N(0)V]$, where N(0) denotes the density of states at the Fermi level and V is the mean matrix element of the interaction energy corresponding to attraction. In the present case $\hbar\omega$ is the energy corresponding to the difference between energy levels of adsorbed atoms which is of the order of 1 eV. Then, considering the characteristic value for $N(0)V \approx 0.4...0.5$ [38], one gets for the superconducting transition temperature $T_c \sim 1000$ K. Thus, superconducting signatures detected in graphite at room temperature, see Figs. 7, 8 and Ref. [2], can easily be accounted for by the Ginsburg's model.

This work has been supported by FAPESP and CNPq Brazilian science agencies.